arXiv:1703.05267v1 [cs.SI] 15 Mar 2017

# Consumers and Curators: Browsing and Voting Patterns on Reddit

Maria Glenski[1], Corey Pennycuff[1], and Tim Weninger[1]

[1] *Department of Computer Science and Engineering, University of Notre Dame*

**Abstract**

As crowd-sourced curation of news and information become the norm, it is important to understand not only how individuals consume information through social news Web sites, but also how they contribute to their ranking systems. In the present work, we introduce and make available a new dataset containing the activity logs that recorded all activity for 309 Reddit users for one year. Using this newly collected data, we present findings that highlight the browsing and voting behavior of the study's participants. We find that most users do not read the article that they vote on, and that, in total, 73% of posts were rated (*i.e.*, upvoted or downvoted) without first viewing the content. We also show evidence of cognitive fatigue in the browsing sessions of users that are most likely to vote.

## 1 Introduction

We frequently rely on online ratings contributed by anonymous users as an important source of information to make decisions about which products to buy, movies to watch, news to read, or even political candidates to support. These online ratings are gradually replacing traditional word-of-mouth communication about an object or idea's quality. The sheer volume of new information being produced and consumed only increases the reliance that individuals place on the ability of anonymous others to curate and sort massive amounts of information. Because of the economic and intrinsic value involved, it is important to understand how individuals consume anonymously-curated information and contribute to the wisdom of the crowd.

*Social news aggregators* like Digg, HackerNews, and Reddit are systems that allow individuals to post news and information for others to view, vote, and comment on. Unlike online social networks such as Facebook and Twitter, social news aggregators typically lack a strong user identity, *i.e.*, users are mostly anonymous and social relationships like friendship or leader-follower do not exist. In the absence of social relationships, social news aggregators rely heavily on user-votes when deciding which content to deliver to the user. In this way, news aggregators allow *information consumers* to also act as *information curators* through their votes.

Social news aggregators represent a stark departure from traditional media outlets in which a news organization, *i.e.*, a handful of television, radio, or newspaper producers, sets the topics and directs the narrative. Socio-digital communities increasingly set the news agenda, cultural trends, and popular narrative of the day [1, 2, 3]. News agencies frequently have segments on "what's trending," entire television shows are devoted to covering happenings on social media, and even live presidential debates select their topics based on popular questions posed to social media Web sites. As this trend continues and grows, it is important to understand not only how individuals consume information that is delivered to them but also how their online behavior shapes the news and information that countless others consume.

When researchers study social media, they study the collective actions of a community by looking at coarse-grained signals of posts, comments, and vote totals. For example, the 90-9-1 rule, representing the proportion of Wikipedia browsers, editors, and creators [4], might be reflected in social news aggregators as the number of browsers, commenters, and posters. Other work in this area has found that user behavior can be easily influenced and manipulated through injections of artificial ratings [5, 6]. This is especially relevant in light of the finding that 59% of bitly-URLs on Twitter are shared without ever being read [7]. Votes, artificial or not, influence the score of a post, which is the primary means of ranking the content. This leads to a "rich get richer" effect where a positive rating directly results in a higher score and therefore a higher ranking; the increased ranking raises the visibility of the post, which provides a better opportunity for other users to view and vote on the post.

Nearly all studies involving the analysis of social media rely exclusively on the information provided by the social media platform itself, typically via API calls or by crawling the Web site. Unfortunately, this means that many questions about browsing behavior, voting habits, and commenting remain unanswered.

**Complete activity logs of Reddit usage.** In contrast to previous studies, our analysis of social media behavior is based on complete activity logs for 309 users of the popular social news aggregator Reddit from August 1, 2015 to July 31, 2016. Study participants were asked to download a browser extension (available for the Firefox and Chrome browsers) that reported usage data of all clicks, pageloads, and votes made within the `reddit.com` domain. Upon installation, the browser extension asked the user to opt-in to the data collection via an informed consent form; the user could modify their consent or uninstall the extension by clicking a small icon that was embedded next to their username in the top-right of the Reddit Web page. This experiment was reviewed and approved by ethics review boards at the University of Notre Dame and the Air Force Office of Scientific Research.

**Recruitment and Sampling Bias.** Study participants were recruited through posts to various subreddits. Participants were required to be a registered Reddit user. Their account must have been created at least a week before installing the browser extension in order to remove the potential for malicious users. We only recorded activity that occurred while the participant was signed in to their Reddit account. Activity that occurred within private or "incognito" browsing modes was not collected. All data was anonymized on the client-side. We did not collect IP addresses of the user, nor did we collect posts or comments made by the user as these could be used to easily de-anonymize the data.

It is important to be cognizant of potential sampling bias. The only way to collect this data is for users to self-select, *i.e.*, opt-in, to the project. Self-selection bias, especially in online polling, for example, will often skew responses in favor of the most motivated group. On the contrary, our system does not ask any questions; it merely observes user behavior. Our recruitment strategy may be affected by undercoverage bias, where certain groups are not included in the sample. Because of the potential for undercoverage bias, we do not intend for this data to be representative of broad or group-based opinion or popularity. Indeed, the opinions and views expressed on Reddit itself are not broadly representative of the general public. Therefore, in this study we only consider *how* registered users browse and vote on Reddit, not *what* they browse and vote on.

This dataset contains about 2 million total interactions and may be used to answer interesting questions apart from those addressed in this paper. To that end, all experimental code is available at `http://dsg.nd.edu/reddit_activity_study` and data will be released upon publication.

In the present work we offer a first look at this newly collected data through two complementary topics:

1. **Information Consumers and Curators.** Because of the critical role that news and information play in everyday life, it is important to understand the dynamics of modern media delivery. The present work is a first look at the private behavior of users of a large social news aggregation Web site. In the first section, we investigate the duality of users in social media, wherein information consumers, *i.e.*, browsers or lurkers, also act as information curators through their votes. With the new data we can begin to answer questions about the variety of information that people browse, and their behavior before and after a vote.

2. **Performance Deterioration in Browsing Sessions.** It is also important to consider the acuity of a user during curation. Human performance is known to diminish after periods of sustained mental effort. The same is probably true in online browsing sessions. Recent work has found that a user's online effort, measured by comment length and writing level, typically declines as a browsing session progresses [8]. However, relatively little is known about how this phenomenon affects voting behavior, which is critical to understand because votes directly determine the content that other people view.

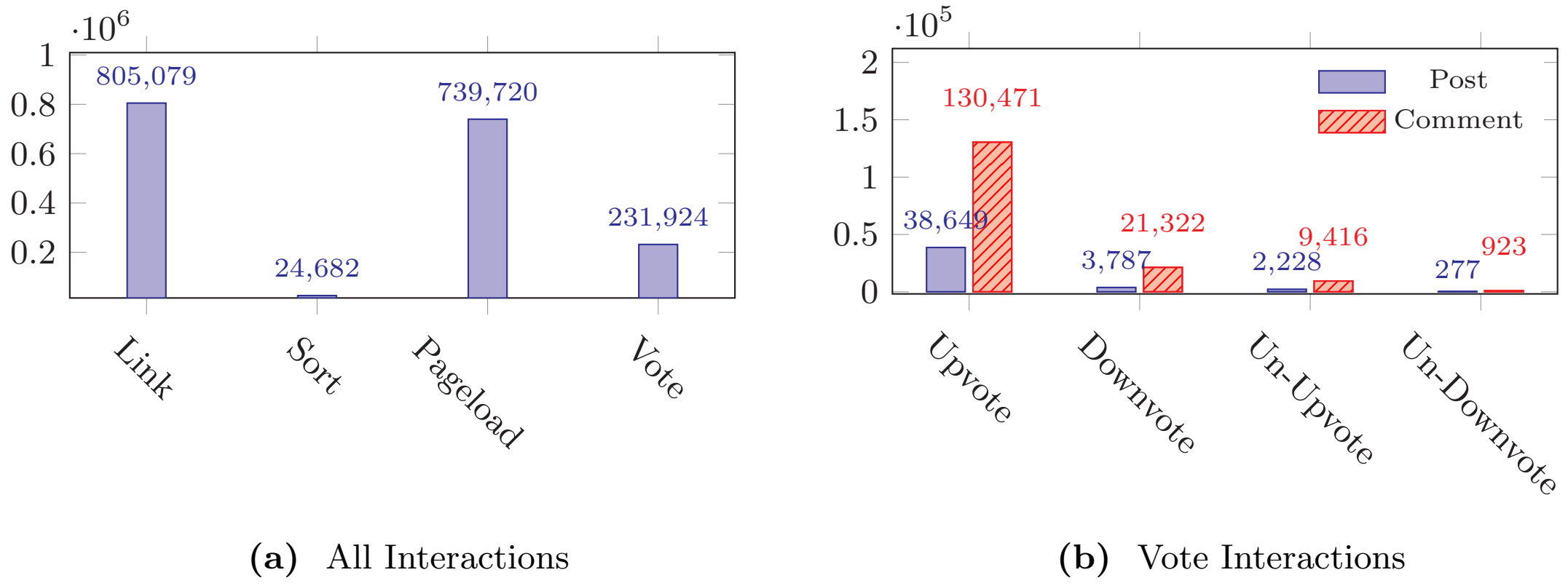

**Figure 1:** Interactions of 309 registered Reddit users between August 2015 and August 2016.

## 2 Consumers and Curators

As a way of introducing the dataset, we begin with some simple aggregate statistics. We have a total of 1,801,405 total interactions from 309 registered Reddit users. Figure 1a breaks down these interactions into their various types. *Link-interactions* occur whenever a user clicks on a content-hyperlink of any kind, including a post's title, a post's comment section, content expanders, etc. Simply put, a link-interaction occurs whenever a user views the content or comments of a post. We find that 89.2% of all link interactions occur from within a *ranked* list, *i.e.*, a link that is ranked amidst other content such as on a user's frontpage or a subreddit. Most of the remainder occur within a comments section, which are *threaded* rather than ranked. Regardless of view, 21.1% of all clicks were on post titles; 20.9% and 16.1% were on image expanders and collapsers respectively; 10.6% of clicks opened the comments section; 5.4% and 3.3% were on text expanders and collapsers respectively; and 4.6% and 3.2% were on video expanders and collapsers respectively.

*Sort-interactions* are those clicks that reorder the ranking of information displayed to the user. The default ordering is `best`, which ranks posts by a time-normalized score function. Re-ordering is relatively rare, occurring in only 3% of all pageloads. When reordered, users choose to sort by `new` 68.1% of the time followed by `top`, `hot`, `rising`, and `controversial` at rates of 14.0%, 9.9%, 6.0%, and 2.0% respectively.

*Pageload-interactions* occur anytime a Web page within the Reddit domain is loaded or refreshed. Thus, a pageload event will duplicate certain types of sort- or link-interactions. For example, a click into a comment section will be recorded as a link-interaction as well as a pageload. Content expanders do not count as a pageload, nor do outbound clicks to non-Reddit Web sites (but outbound clicks do count as link-clicks). We find that the Reddit `frontpage` was loaded in 23.6% of all pageloads, followed by `r/all` with 4.8% of all pageloads, further followed by a litany of subreddits.

*Vote-interactions* are those clicks that upvote, downvote, un-upvote or un-downvote a post or a comment. In order to un-vote, a user must have previously cast an initial vote; we do not substract un-votes from the vote totals. Figure 1b shows the breakdown of votes in our dataset. Interestingly, we find that there are 357% more votes on comments than there are on posts, perhaps due to the relative abundance of comments.

It is important to note that we only intend for these aggregate statistics to describe the size and scope of our data collection. It is likely that the interactions of a handful of power users are able to skew these results. A better way to analyze consumer and curation behavior on Reddit is to investigate interactions from a user perspective. Thus, throughout the remainder of this section we investigate social media consumption and curation interactions by plotting the distributions of users within each task.

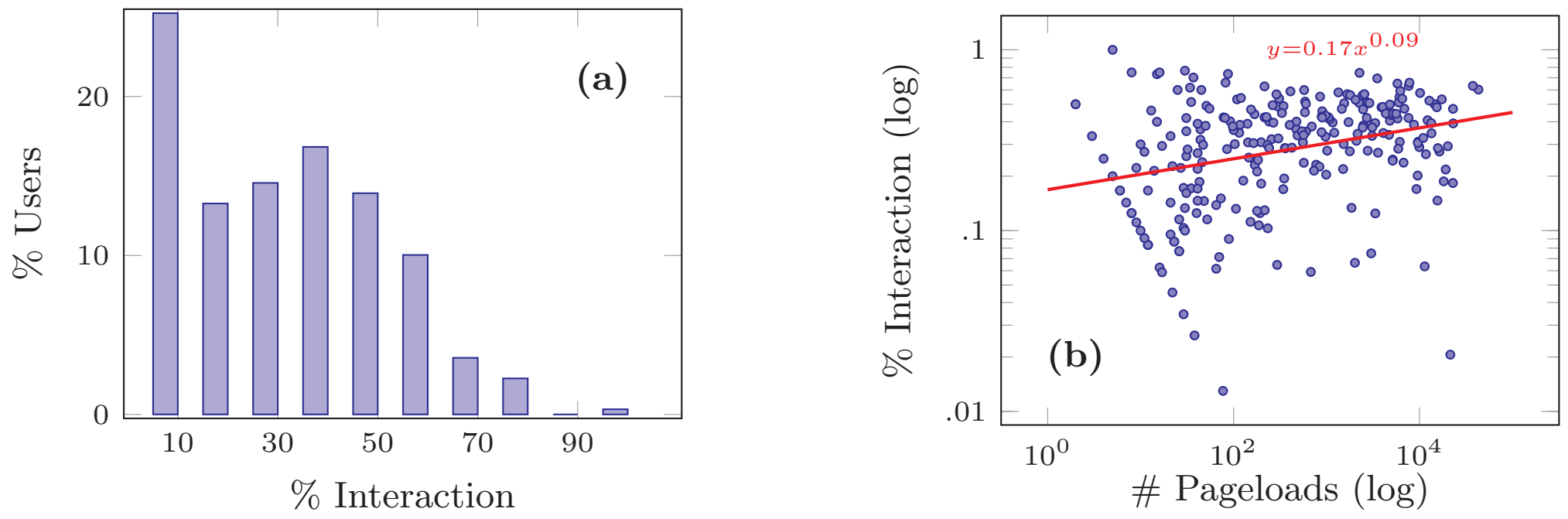

**Figure 2:** Headline browsing tendencies. **(a)** shows a histogram of percentage of users as a function of their tendency to interact with each pageload. The typical user interacts with a loaded page about half the time. **(b)** plots an un-grouped scatterplot of the same data. The scatterplot indicates that the number of pageloads is weakly correlated with the user's tendency to interact the content ($R^2 = 0.09$, $p < 0.01$) .

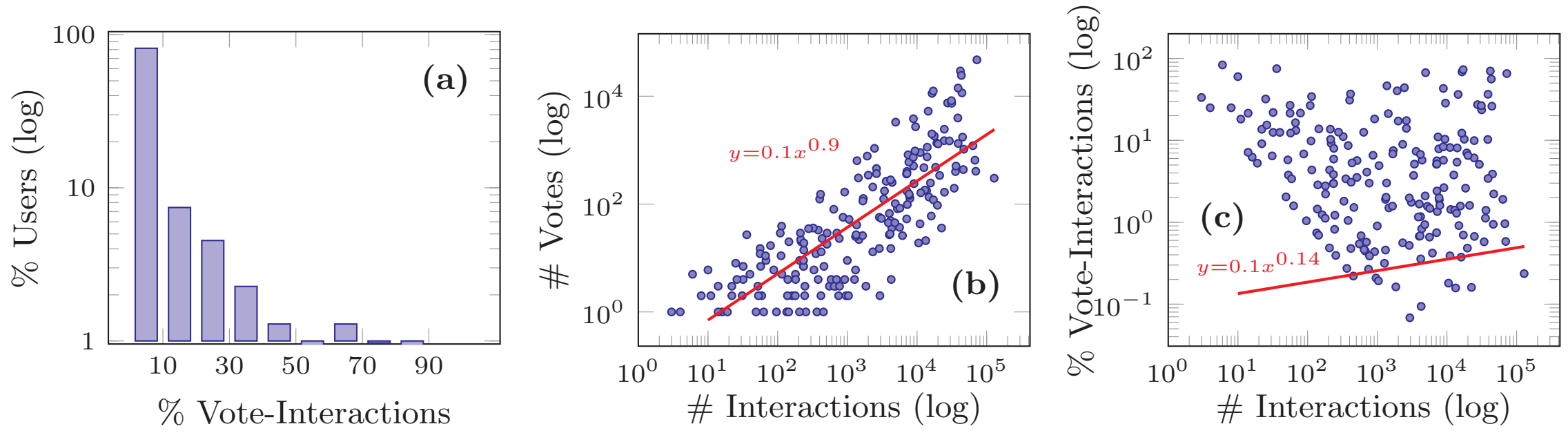

**Figure 3:** Voting tendencies. **(a)** plots a histogram of the number of users as a function of their tendency to vote. The scatterplot in **(b)** indicates a correlation between the number of a user's interactions that were vote-interactions and the total number of a user's interactions ($R^2 = 0.64$, $p < 0.01$). The scatterplot in **(c)** indicates a weak correlation between the proportion of a user's interactions that were vote-interactions and the total number of a user's interactions ($R^2 = 0.04$, $p < 0.01$).

## 2.1 Headline Browsers

Previous studies frequently show that the vast majority of Web users do not contribute content [9, 10]. However, these studies all tend to assume the non-contributors still generally browse the content of a post. The problem with this assumption is that social media aggregators do not show the content initially; instead, they present the user with a ranked list of short headlines, which are usually carefully crafted to encourage the reader to click to see more. For our first task, we ask: How often do users browse headlines without clicking through to see the content?

For this task, we want to find the number of users who loaded a page on Reddit, but did not interact with the content on the page at all, *i.e.*, the user only browsed the headlines. We define a *content-interaction* as any interaction that opens the post's content either via content-expander, click-to-comments, click-to-external-URL, etc. Although we cannot determine what a user read but did not click, it is reasonable to assume that nearly all pageloads were purposeful, and that the user read at least part of the loaded Web page.

Figure 2 shows the interaction rates as a histogram (left) and as a scatterplot (right). We find that 25% of participants had content-interactions with less than 10% of their pageloads. The data shows that most study participants were *headline browsers*. Specifically, 84% of participants interacted with content in less than 50% of their pageloads, and the vast majority (94%) of participants in less than 60% of their pageloads. Because of the cognitive effort required to interact with each post, we expected to see a negative correlation between the number of pageloads and interaction percentage. However, the scatterplot in Fig. 2 (right) shows that a user's propensity to interact with a page is only slightly correlated with the number of pageloads ($R^2 = 0.09$, $p < 0.01$).

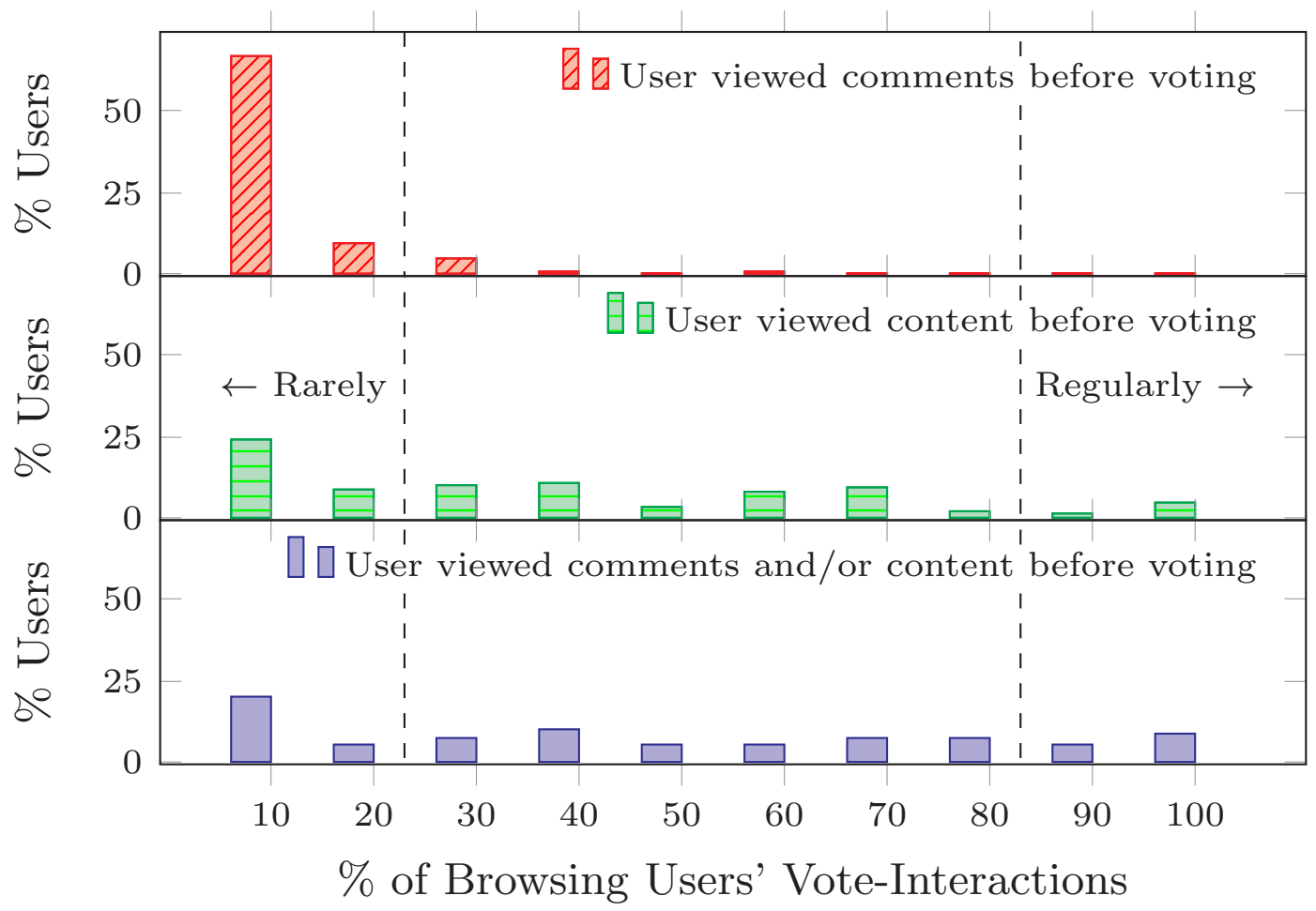

**Figure 4:** User behavior before voting on posts. Bar plots illustrate the likelihood that users vote on a post after viewing its comments section (top plot), content (middle plot), or either (bottom plot). The bottom plot shows that nearly one-third of voters rarely (*i.e.*, < 20% of the time) view the content or comments of a post before casting a vote.

## 2.2 Vote Behavior

Unlike *browsing interactions*, *i.e.*, viewing the post content or comments section, described above, *voting interactions* are critical to the functioning of social media aggregators like Reddit because they provide explicit feedback to the ranking system, which uses the ratings to reorder the content. Here we look specifically at these vote-interactions to better understand voting behavior.

Like previous studies which have found that most users do not vote on the majority of posts [11], we show in Fig. 3 (left) that many of our participants do not vote often or at all. About 50% of users vote in less than 1% of their interactions. While the middle scatterplot in Fig. 3 shows that a user's number of vote-interactions is positively correlated with the user's total number of interactions ($R^2 = 0.64$, $p < 0.01$) when fitting a power regression, the scatterplot on right shows that a user's tendency to vote (the proportion of their interactions that were votes, denoted % Vote-Interactions) is only very weakly positively correlated with the user's total number of interactions ($R^2 = 0.04$, $p < 0.01$). As a user's total number of interact increases, they tend to have a larger number of vote-interactions but not an increased tendency to vote.

When users do vote, we ask: How often do users actually read the content of the post before they vote on it? To answer this question, we examined all of the posts that were voted on and for each user that cast a vote, we counted the number of times that the user viewed the content or comments section of the post before voting. The distribution of these counts, as a proportion of the user's total number of vote-interactions (denoted % of User's Vote-Interactions) for all participants with vote-interactions is plotted in Fig. 4. The bottom bar plot shows that almost one-third (31%) of all participants that voted only rarely (*i.e.*, < 20% of the time) viewed a post's content and/or comments before voting and thus make their decision based on the headline alone. Conversely, about 17% of all participants regularly (*i.e.*, ≥ 80% of the time) viewed a post's content or comments before voting. The remaining users are spread relatively equally between the two extremes of rarely and regularly browsing before voting.

Similarly, the top plot in Fig. 4 shows the distribution of users that view the comments section of a post before they vote, and the middle plot shows the distribution of users that view the content of a post before they vote. These plots are not mutually exclusive, *e.g.*, a user may view both the content and comments section before voting. We find that viewing the comments rarely precedes a vote, but the likelihood that a user views the content of a post before voting has a more even distribution.

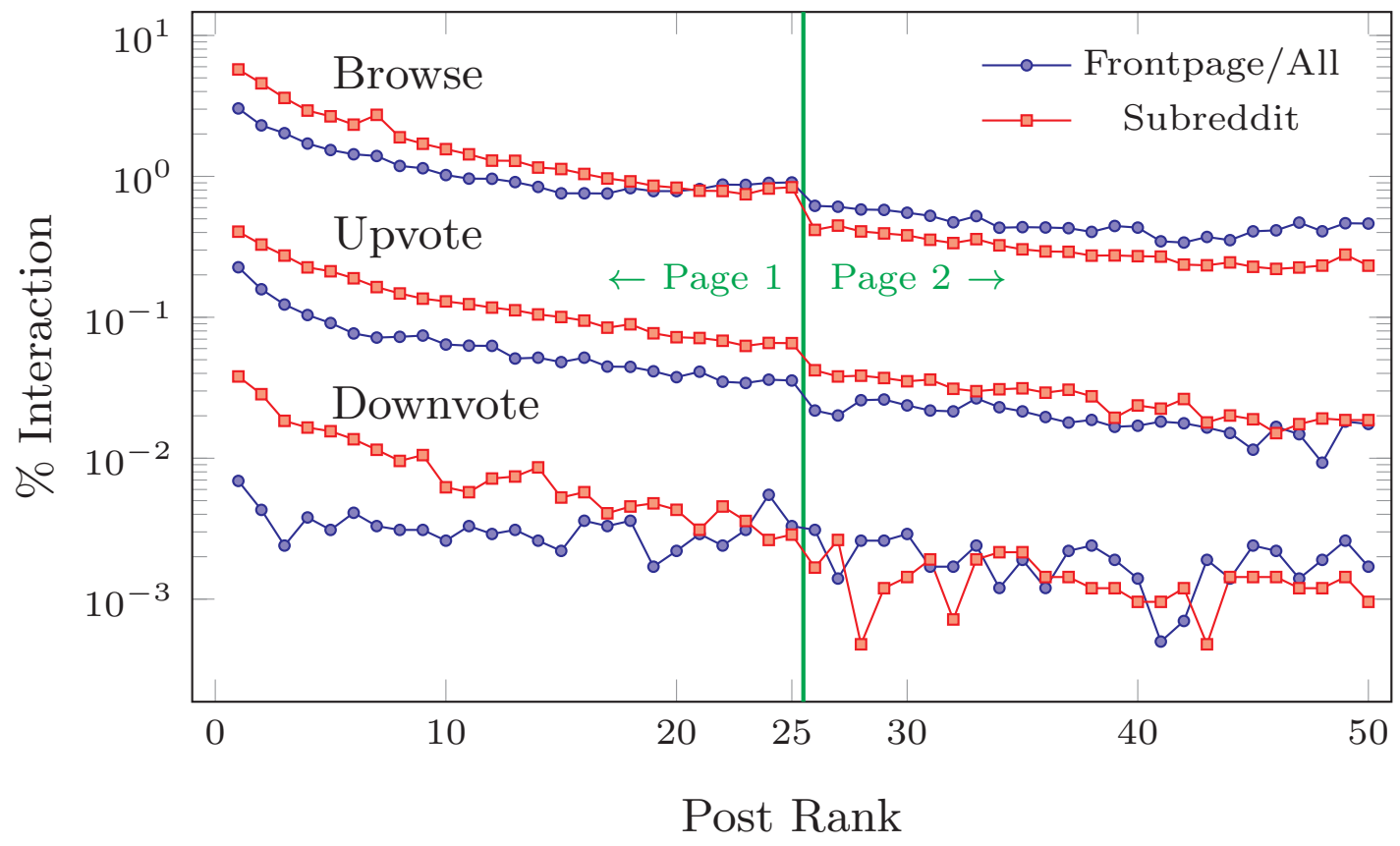

**Figure 5:** Effects of rank on tendencies to browse, upvote, or downvote posts from a user's frontpage (or /r/all) or a specific subreddit.

Browsing the content of a post is certainly a more demanding task than not browsing. So we expected to see that frequent voters are more likely to vote without viewing the post's content. However, contrary to our expectations, we found that there was no correlation ($R^2 = 0.06$) between a user's total number of votes and a user's tendency to vote without browsing (not illustrated). In total, out of 41,540 posts that were voted on, 30,421 (*i.e.*, 73%) were rated without first viewing the content or comments of the post.

## 2.3 Position Bias

Another well studied phenomenon found in social news aggregators is *position bias*, wherein people tend to pay more attention to items at the top of a list than those at the bottom [12, 13]. As a consequence of position bias, the presentation order has a strong effect on the choices that people make [14, 15]. This is especially true for Web search results [16], and has even been found to affect the answers that people select when answering multiple choice questions [17].

Here we describe position bias on Reddit and also show how position affects voting behavior. Before we begin, it is important to note that each Reddit post must be submitted to a *subreddit*. A subreddit is a topic-specific community (*e.g.*, /r/news, /r/ama, /r/machinelearning) with its own set of relevancy rules, guidelines and mores. If a registered user subscribes to some subreddit, then the posts submitted to that subreddit will appear on the user's *frontpage* in an order computed by one of Reddit's ranking systems (*e.g.*, best, top, rising, new). A user can also view the posts submitted to a specific subreddit by navigating, searching, or typing the subreddit's name. Alternatively, a user can view the posts from all subreddits, regardless of subscription status, by visiting the special /r/all subreddit. The choice of view, either the frontpage/all or a specific subreddit, drastically changes the scope of the posts that the user sees, and may also change how the user interacts with the content.

Figure 5 shows the effect that rank has on the likelihood that study participants browsed (*i.e.*, viewed the content or comments section), cast an upvote, or cast a downvote on a post. The interaction percentages are divided into interactions that occurred from a mixed-subreddit view (*i.e.*, the frontpage or /r/all) or from a specific subreddit. As we observed earlier, the number of browse interactions, upvotes, and downvotes differ by an order of magnitude. In Fig. 5 we plot the percentage of each interaction at each rank as a percentage of the total interactions within each view. For example, the top-left point shows that 7.9% of all clicks in a subreddit were to browse the content or comments of the top-ranked post.

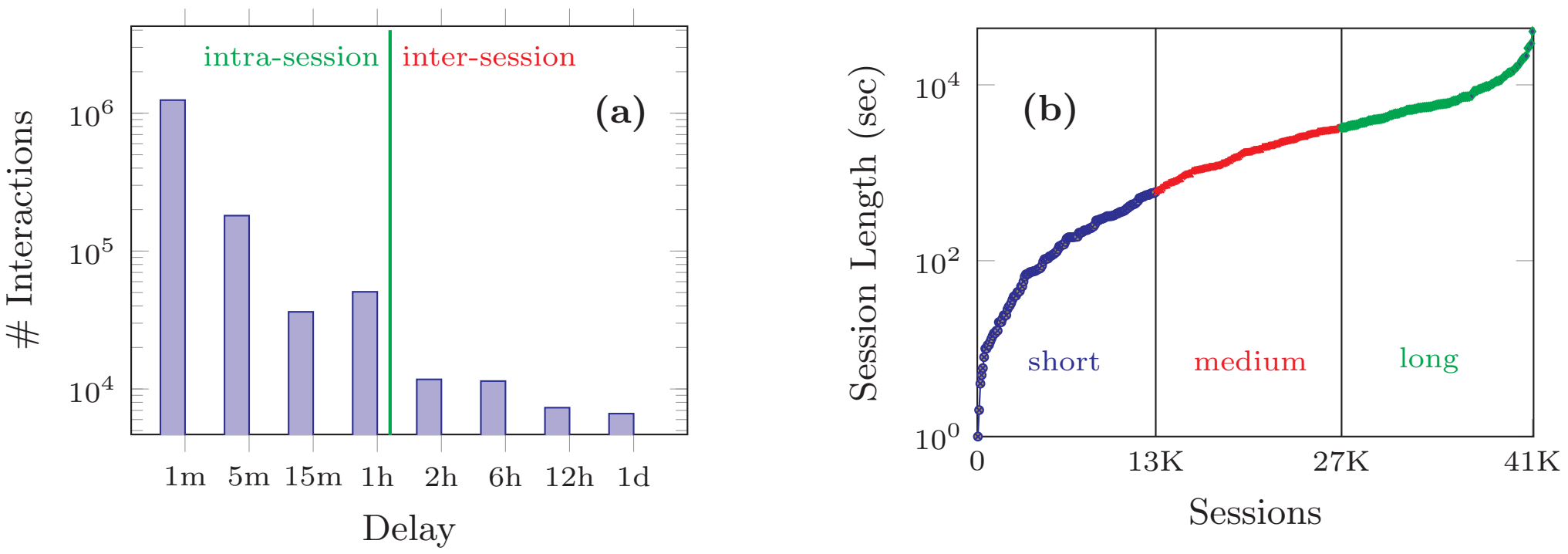

**Figure 6:** Sessionizing Heuristic. A histogram of inactivity delays between consecutive interactions in **(a)** shows the uptick at 1 hour that signals the boundary between intra- and inter-session delays indicating the use of a session inactivity ceiling of 1 hour. The distribution of resulting session lengths in **(b)** shows where we partitioned the sessions into short, medium, and long sessions.

Position bias is clearly evident in our results: a user is about 4 times more likely to interact with the top post than the 10th post. Line plots of browsing behavior show an uptick in interaction likelihood just before rank 25. This increase has been previously attributed to ‘contrarians’, who navigate the list backwards [18, 15]. A large decrease in browsing likelihood is found between rank 25 and 26, which corresponds to Reddit’s default pagination break.

Across all interaction types, posts are more likely to be browsed, upvoted, and downvoted when viewed from within their specific subreddit, than when viewed from the front page. After the page-break, the gap between views narrows and even flips in the case of browsing interactions. This indicates that users are more willing to explore further down the ranked list when presented with a more diverse set of posts.

# 3 Browsing Sessions

In this section, we present several experiments that highlight the effects of user behavior over the course of a browsing session. For these tasks, we first need to identify the browsing sessions for each user. In the collected data, a timestamp is recorded for each interaction and a *browsing session* is loosely defined as a period of regular activity. In a recent study, Halfaker *et al* presented a methodology for identifying clusters of user activity by plotting the histogram of the time elapsed between interactions. They argue that the appearance of regularity within delays implies a good sessionizing heuristic. The results of Halfaker *et al* as well as sessionizing results for delays between comments by Singer *et al* found that a delay threshold of about 1 hour is an appropriate threshold [8, 19]. In other words, these studies suggest that if a user is inactive for more than 1 hour, then that user’s session has ended and the next activity will begin a new browsing session.

Using the same methodology, the inter-activity delays of our data is plotted in Fig. 6 (left). We can clearly see an uptick in the number of delays that occur in the time-bin prior to one hour. These results are in alignment with the previous studies, so we confidently define a browsing session for a particular user as a contiguous block of activity with delays of at most one hour. This created 41,385 unique browsing sessions across all 309 users for a mean-average of 133 sessions per user over the year.

Not all users are equally active, and the activity of each user in a session varies significantly. For instance, although the mean-average session length is about 53 minutes (median: 21 minutes), Fig. 6 (on right) shows that many browsing sessions last for only one-click, while the longest session lasted for several hours. In our analysis, we partitioned the sessions into equal-sized bins of short, medium, and long, resulting in session length groupings of <3 min, 3min-53min, and >53 min respectively. We then classified users as being short-, medium-, or long-browsing users based on the session length of the majority of the user’s sessions. Figure 7 summarizes the distribution of users and sessions across

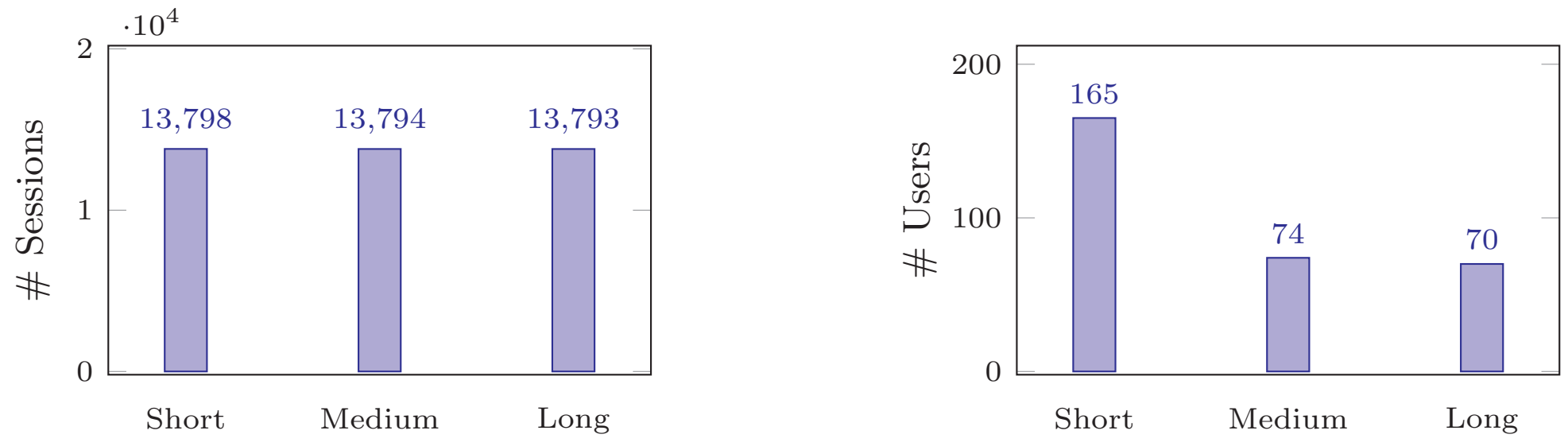

**(a)** Session Length

**(b)** User Type

**Figure 7:** Distributions across categories based on their sessions by length **(a)** and by users **(b)**.

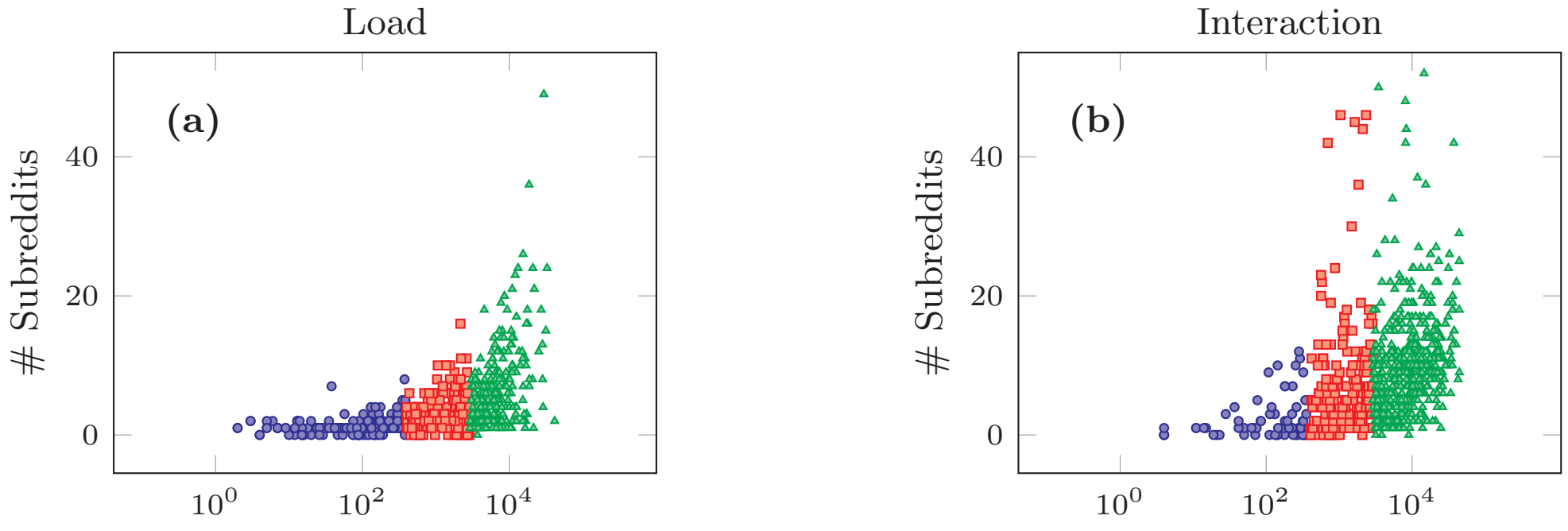

**Figure 8:** Community Variety. Distinct subreddits loaded (left) and interacted with (right) during a session by session length. Despite being highly skewed, the mean-average number of subreddits loaded are 1.4, 2.5, and 4.8 for short-, medium-, and long- browsing users, respectively, and even lower 1.0, 1.3, and 1.5, respectively, for subreddits users actively interacted with.

types. We find that most users in our dataset had primarily short-browsing sessions, *i.e.*, the plurality of their sessions lasted less than three minutes, and nearly equal numbers of medium-browsers and long-browsers.

## 3.1 Community Variety

Despite the potential for social news and social networking sites to expose users to diverse content and new perspectives, many recent studies show the opposite effect. Through their friendships, filters, subscriptions, etc., users frequently filter out those ideas and opinions that clash with their own [20, 21]. This has led to the rise of the "filter bubble" and the "echo chamber" wherein users only view what they agree with [22], and has been associated with the adoption of more extreme views over time [23] and even misconception of factual information [24]. We expect to find a similar echo chamber on Reddit because a user's frontpage is populated from only those subreddits, which are topic-based communities, to which the user subscribes. Here, we take a first look at the variety of the communities browsed on Reddit within browsing sessions.

Unfortunately, it is difficult to computationally discern the ideological leanings for each of the 5,235 distinct subreddits in our data set to perform content-based analysis of variety. Instead, we look at the number of distinct subreddits that a user loads as a measure of the variety of communities that a user views content from. While the content within a single community may cover a narrow or diverse range of topics, content is contributed and curated by the user-base of that community. Since we have seen above, in Fig. 2, that typical users have a low tendency to interact with each pageload, we use the number of distinct subreddits interacted as another estimate at the community variety of a browsing session, in this case as a proxy of the variety of communities the user more actively engages with. Figure 8 shows, as expected, that each session-type and user-type exhibits different browsing behavior. It is reasonable to expect that the number of subreddits found within a short-browsing session will be smaller than the number of subreddits found within long browsing sessions simply because longer sessions have a greater opportunity to view more content or communities.

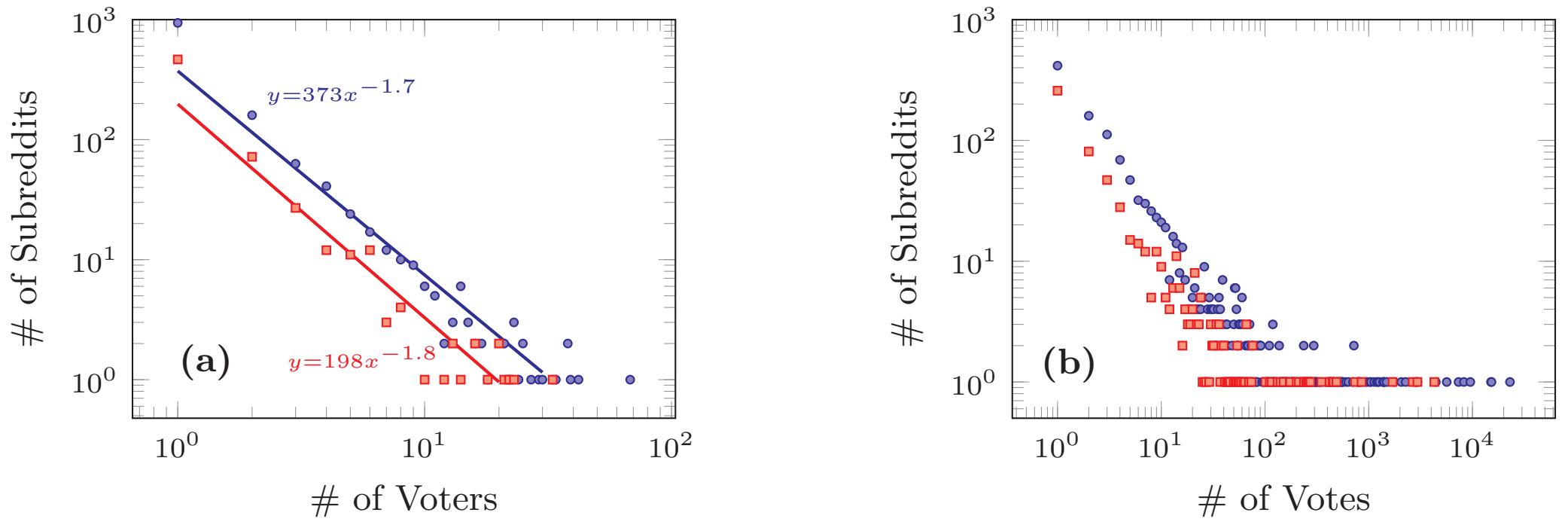

**Figure 9:** Voting Variety. Number of subreddits by number of voting users, illustrated by **(a)**, exhibits a power law distribution. Number of subreddits by total number of votes, illustrated by **(b)** also exhibits a power law distribution with a much longer tail. Blue circles and red squares indicate upvotes and downvotes respectively.

Despite the distributions illustrated in Fig. 8 being highly skewed left, we find a surprising lack of variety when we look at the mean-averages. We find that users navigate to a mean-average of 1.4, 2.5, and 4.8 distinct subreddits for short-, medium-, and long-browsing user-types respectively. Despite the number of subreddit pageloads, we further find that users only interact with posts from 1.0, 1.3, and 1.5 subreddits from short-, medium-, and long-browsing users on average. In other words, most users only click or vote on a very specific set of posts, even in long-sessions. The discrepancy between subreddit-pageloads and interactions can be explained as headline browsing behavior; that is, even if a user views a variety of subreddits, it is unlikely that the user will actually click past the headline. If plotted as a function of session-lengths rather than user-type, Fig. 8 illustrates the same lack of community variety with nearly identical the mean-averages.

The dearth of community variety also occurs across browsing sessions. Despite collecting 1,018 subscribe-events and 437 unsubscribe-events over the year-long data collection period, the median user subscribed to 0 new subreddits. In fact, all of the subscription events can be attributed to only 109 unique users (104 subscribed, 44 unsubscribed), and from those 104 subscribers only 26 subscribed to 10 or more new subreddits. In summary, although many of our study participants spent a lot of time interacting with Reddit, we find that most users, even highly active users, perform most of their activity within a handful of subreddits.

Because of the demonstrated lack of browsing variety, we also suspect a lack of voting variety. Figure 9 (on left), which illustrates the distribution of voters to subreddits, shows that the majority of subreddits in our collection have only one voting user and that the distribution resembles a power law distribution. Figure 9 (on right) illustrates the distribution of votes to subreddits. The distribution of subreddits to votes (on right) has an even longer tail than the distribution of subreddits to voters (on left) because a single user may interact with many posts and/or comments within a single subreddit thereby severely skewing the results. In both plots the finding is the same: most subreddits receive very few votes from very few voters; only a handful of subreddits receive the majority of the votes.

## 3.2 Performance Deterioration

It is widely recognized that human performance tends to deteriorate during periods of sustained mental or physical effort. Physical fatigue is most often associated with prolonged exercise or exertion where the physical performance of an individual diminishes over time. Similarly, researchers have found evidence of cognitive fatigue where an individuals ability to complete cognitive tasks diminishes over time. Understanding the processes behind performance deterioration is critical in many enterprises where individuals must maintain high performance over long periods. Many complicated and delicate surgeries, for example, may take hours to complete, yet still require high cognitive performance [25]. Many other high-stakes tasks, such as flying and driving, are equally prone to performance deterioration [26]. Cognitive fatigue has also been studied in test taking, where individuals must have consistent performance on cognitively demanding tasks that take long periods [27].

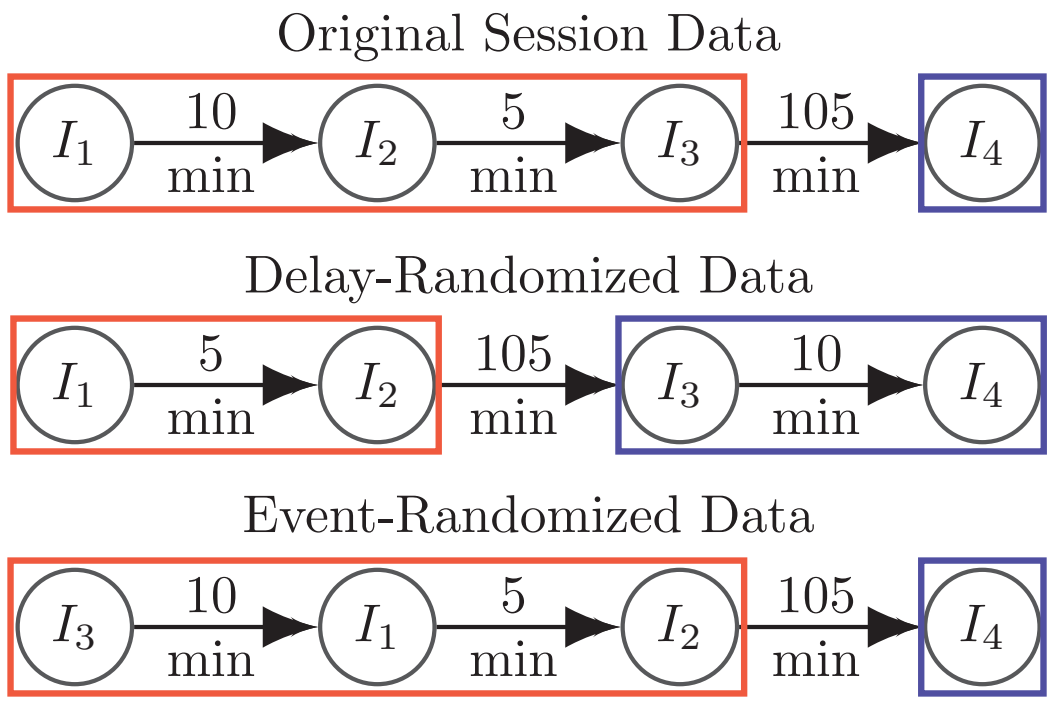

**Figure 10:** Examples of session partitioning for the original, delay-randomized, and event-randomized datasets. Circles represent interactions $I_j$, arrows represent inactivity delays between consecutive interactions $I_j$ and $I_{j+1}$, and boxes distinguish between sessions. The first row illustrates an example of how data is partitioned into sessions in the original dataset; the second row shows how sessions change when delays are randomly shuffled; and the third row shows how the sessions change when the events are shuffled.

As society increasingly relies on the ratings of users to curate online news and information, it is important that we understand whether and how cognitive fatigue affects online user behavior. In this section, we explore this topic by looking at user behavior over the course of a browsing session.

To understand how user behavior changes over the course of a session, we must first develop comparative baselines. To that end, we generated two randomized session datasets from the observed data that we label as *delay-randomized* and *event-randomized* sessions [8]. In the delay-randomized dataset, we randomly shuffle the ordering of the delays between interactions while holding the ordering of interaction events constant. For the event-randomized dataset, we randomly shuffle the ordering of the interaction events while holding the delay between interactions constant. We illustrate how sessions differ between the original, delay-randomized, and event-randomized datasets in Fig. 10[1]
.

The interactions of each browsing session are placed into decile bins, where the first 10% of interactions [0,10] are placed in the same bin, the second 10% (10,20] are placed in the same bin, and so on. Decile binning provides an opportunity to analyze aggregate statistics about the behavior of participants as they progress through their session. As in previous analysis, we separate participants by their typical session lengths into short-browsing, medium-browsing, and long-browsing users using the same criteria as in the previous section. Using this methodology, we can begin to ask questions about the activities performed by users over the course of their browsing session.

We first look at the ranks of posts with which a user interacts. Reddit orders content in various ways. Generally, and by default, posts are ranked by the number of upvotes minus the number of downvotes normalized by the time since submission. Based on the interaction percentages by post rank illustrated in Fig. 5, we expect that readers begin their browsing session by reading and interacting with the top ranked posts on the frontpage, followed by lower ranked posts. From the subreddit results in Fig. 8, we find that many users, especially medium- and long-browsing users, often navigate to subreddits and continue to browse topical content.

Based on these initial findings, we expect that the median rank of posts that are viewed or voted on will rise as the user session progresses. To test this hypothesis, Fig. 11 (on left) shows that the median post rank is statistically correlated with the session decile for all user-types. As expected, short-browsing users have a higher correlation than medium- and long-browsing users because longer browsing-users are more likely to view more subreddits, each of which start their numbering at 1. These results are in contrast to the randomized baselines drawn in Fig. 11 (center) and Fig. 11 (on right), which mostly show non-significant correlations.

[1]Figure adapted from [8] with permission.

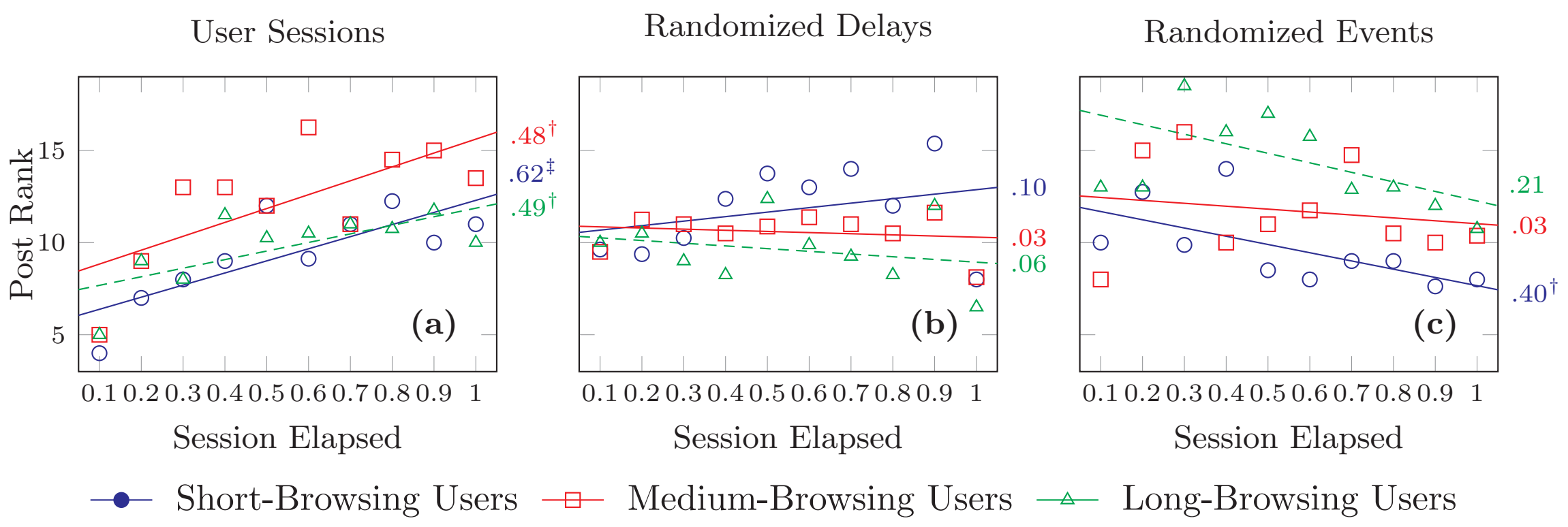

**Figure 11:** Median post rank as a function of the session decile for short-, medium-, and long-browsing users. Linear regression lines are plotted, and the correlation of determination $R^2$ value is shown to the right of each regression line. $^\dagger$ and $^\ddagger$ symbols represent statistically significant effects with $p < 0.05$ and $p < 0.01$ respectively. The user sessions plot **(a)** shows that the median ranks of browsed posts tends to increase for all types of users. As expected, baseline plots illustrating session shuffled by randomized delays **(b)** and randomized events **(c)** show little, if any, correlation.

**Table 1:** Slope ($m$) and effect sizes ($R^2$) of various factors as a function of session progress. Regression plots for post Rank viewing are illustrated in Fig. 11 (on left), and regression plots for SRNΔ Upvote (SRNΔ^) are illustrated in Fig. 12 (on left). $^\dagger$ and $^\ddagger$ symbols represent statistically significant correlation with $p < 0.05$ and $p < 0.01$ respectively.

| | **User-Type** | | | | | |
|---|---|---|---|---|---|---|
| | **Short** | | **Medium** | | **Long** | |
| | $m$ | $R^2$ | $m$ | $R^2$ | $m$ | $R^2$ |
| **Post Rank** | | | | | | |
| **View** | (+) | .62$^\ddagger$ | (+) | .48$^\dagger$ | (+) | .49$^\dagger$ |
| **Upvote** | (+) | .24 | (+) | .22 | (+) | .33 |
| **Downvote** | (+) | .01 | (+) | .13 | (0) | .00 |
| **Post Score** | | | | | | |
| **Time of View** | (-) | .42$^\dagger$ | (-) | .01 | (-) | .28 |
| **Final Score** | (-) | .44$^\dagger$ | (-) | .20 | (-) | .57$^\dagger$ |
| **SRNΔ** | | | | | | |
| **Upvote** | (-) | .09 | (+) | .04 | (-) | .78$^\ddagger$ |
| **Downvote** | (-) | .07 | (-) | .42$^\dagger$ | (-) | .00 |

We performed the same regression analysis on several other performance factors. Rather than visually illustrating each graph, we show a a summary of the regression results in Tab. 1. The coefficient of determination ($R^2$) values are tabulated along with signs of their slopes for each user browsing type. The top row repeats the $R^2$ values and tests of statistical significance results from Fig. 11 (on left). The rank of upvote and downvote interactions is not highly correlated with session progress, meaning that although users are likely to browse lower-ranked items as their session progresses, they are not more likely to vote on them.

Unlike a post's rank, which is typically a value between 1 (the top ranked post) and 25 (the bottom of the first page), the score of a post is highly skewed and varies significantly depending on the subreddit to which it was submitted. We ignore these confounding factors, for now, and tabulate their $R^2$ values in the middle row of Tab. 1. A post's score at the time it was viewed is significantly correlated with session progress in short-browsing users, but not for medium- and long-browsing users. Generally, these results have the same interpretation as the post rank results: Short-browsers typically view posts with high scores at the beginning of their session and posts with progressively lower scores as their session continues. We believe that non-correlation in the case of medium- or long-browsing users is because medium and long-browsing users are more likely to visit multiple subreddits (as indicated in Fig. 8), which rank their posts separately.

Two weeks after the end of the data collection period, we crawled each of the recorded posts to determine their "final" score. Like the score at time of view, Tab. 1 shows that the final score of each post that a user viewed is negatively correlated with session progress in short- and long-browsing users.

Although informative as to the nature of social media browsing, these results say little about whether users have any cognitive fatigue. In the next experiment we ask: Is *rating performance* correlated with session progress?

To answer this question we first need to define rating performance in terms of social media sessions. The change in a post's score from the time of view to the final score $\Delta\text{Score}_{I,F}$ is a good starting point because this metric gives an indication of how much the post changed after it was interacted with (*i.e.*, upvoted or downvoted). Positive values for this metric will indicate that a post's score increased after the user's vote; the higher the increase, the more predictive power that the user displayed. For example, a user who interacts with a popular post after it was identified as popular by others (*i.e.*received a large proportion of its total votes) would have a lower $\Delta\text{Score}_{I,F}$ than a user that interacted with the same post earlier when there was a weaker signal of popularity, thus indicating more predictive power.

Unfortunately, this metric alone is highly dependent on the size of the subreddit to which the post was submitted. For example, in smaller subreddits a score of 10 or 20 is considered to be a high score whereas in large subreddits a score of 10 or 20 is actually a very low score. To adjust for subreddit size effects, we normalize $\Delta\text{Score}_{I,F}$ by the absolute value of the median final score of all collected posts within a subreddit. To ensure a representative sample size, we only consider subreddits that have 50 or more posts collected in our dataset. We abbreviate this function as $\text{avg}_{\text{SR}}(\text{Score})$, and the final upvote performance metric is:

$$\frac{\Delta\text{Score}_{I,F} + 0.5}{|\text{avg}_{\text{SR}}(\text{Score})| + 0.5},$$

where +0.5 is added as Laplace smoothing to avoid division by zero. We call this metric the subreddit normalized change in score (SRN$\Delta$), which can be in response to an upvote (SRN$\Delta^{\wedge}$) or a downvote (SRN$\Delta^{\vee}$).

SRN$\Delta$ has a couple of important properties. First, by normalizing by the average of all posts' final scores within each subreddit, this metric partially represents the power of a single vote. Second, SRN$\Delta$ indicates the change as a proportion of the average score of posts within the subreddit, where positive values indicate an increase in the score of the post between the time of a user's interaction and the end of voting.

Figure 12 shows the results of the upvote performance analysis. We do not find any correlation among the variables in the randomized session plots. Within the original user sessions, we find statistically significant negative correlations between rating performance and session progress in the long-browsing user sessions, that is, scores of posts that long-browsing users vote on early in their session tend to increase more than posts that they vote on later in their session.

Because these experiments are not randomized, we cannot say for certain whether the diminishing rating performance indicates a decay in the user's influence, predictability, or some other characteristic. Earlier work in this area has found significant influence effects [5, 6, 28] on a user's vote due to ranking bias [13], but we make no such claim here.

There may exist other confounding factors that explain the deterioration in rating performance. To address this potential, we performed analysis of covariance (ANCOVA) tests for each user-type (*i.e.*, short-, medium-, long-browsers). The ANCOVA test is like the standard analysis of variance (ANOVA) test except that ANCOVA compares the response variable (*e.g.*, SRN$\Delta$), with an independent variable (session progress) and a factor (session ordering) [29].

By comparing the regressions of the original session data against the respective regressions of the delay-randomized and event-randomized sessions, the ANCOVA test can determine if the differences in regressions are statistically different from the random baselines. Results for each experiment are tabulated in Tab. 2. Short- and medium-browsing users are not significantly different from the random baselines, but the long-browsing users have a significant negative decay.

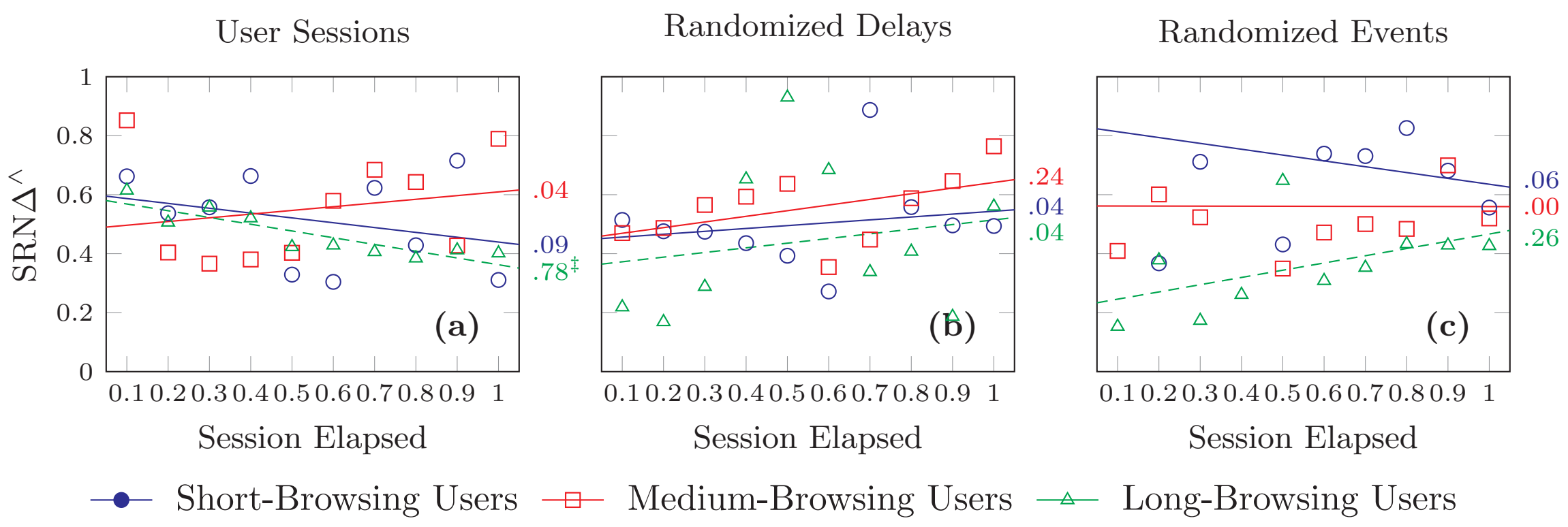

**Figure 12:** Median subreddit-normalized change in score for upvoted posts (SRNΔ^) as a function of the session decile for short-, medium-, and long-browsing users. Linear regression lines are plotted, and the correlation of determination $R^2$ value is shown to the right of each regression line. † and ‡ symbols represent statistically significant effects with $p < 0.05$ and $p < 0.01$ respectively. The user sessions plot **(a)** shows how the median SRNΔ^ changes as a session progresses compared to randomized delays **(b)** and randomized events **(c)**. the median SRNΔ of browsed posts tends to diminish for all types of users, but only long-browsing users show statistically significant correlation.

We experimented with several alternatives to the normalization used in SRNΔ including: (1) using the mean score of a subreddit, rather than the median; and (2) normalizing by the average change in score for each subreddit avg($\Delta \text{Score}_{I,F}$), rather than the average final score. Various combinations of these alternatives created small changes in the regression and ANCOVA analysis, but did not alter the results of significance tests.

# 4 Discussion and Conclusions

The present work presents a user study of browsing and voting behavior on the social news aggregator Reddit. We find that the vast majority of participants are headline browsers, who only view the summary headlines without any further interaction to view the content or read the comments. This shallow browsing behavior is also evident in voting interactions where 73% of posts were voted on without first viewing the post's content and nearly a third of voters almost always (*i.e.*, <20% of the time) vote without browsing a post's content or its comments. Voting dictates the visibility of individual posts and directly determines the content that is presented to others. Our results indicate that voting is probably heavily dependent on the quality of a post's title or its headline rather than the content itself [30]. When combined with the clear evidence of position bias in our dataset, this dependence raises concerns about the manner by which platforms rank content through voting using an assumption that votes correspond with the quality of the content itself.

Interestingly we did not find evidence to link browsing behavior before voting or the effect of position bias with cognitive fatigue, with no statistically significant correlation between the number of votes a user casts and the effort expended before the vote, *i.e.*, whether a user browsed the content or comments of a post before voting. In fact, when we investigated performance deterioration within sessions, we found more evidence of deterioration for long-browsing users than short- or medium-browsing users. While there was some statistically significant increase in the rank of posts interacted with by short- or medium-browsing users when compared to our events-randomized baseline, there was no statistically significant decline in rating performance. Long-browsing users, however, had a statistically significant decline in the predictive power of their upvotes, which we approximated with our SRNΔ^ metric, as their session progressed. This decline could be indicative of a loss of predictive power or accuracy because of cognitive fatigue in long-browsing users.

**Table 2:** Effect sizes ($R^2$) for analysis of covariance (ANCOVA) tests that compare regressions on user sessions against delay-randomized and event-randomized sessions. $^\circ$, $^\dagger$, and $^\ddagger$ symbols represent results of significance tests with $p \geq 0.05$, $p < 0.05$, and $p < 0.01$ respectively between the observed behavior at the delay-randomized (D) and event-randomized (E) sessions.

| | **User-Type** | | | | | | | | |
|---|---|---|---|---|---|---|---|---|---|
| | **Short** | | | **Medium** | | | **Long** | | |
| | $R^2$ | D | E | $R^2$ | D | E | $R^2$ | D | E |
| **Post Rank** | | | | | | | | | |
| **View** | .42 | $^\circ$ | $^\ddagger$ | .25 | $^\circ$ | $^\circ$ | .37 | $^\ddagger$ | $^\ddagger$ |
| **Upvote** | .15 | $^\circ$ | $^\circ$ | .20 | $^\circ$ | $^\dagger$ | .37 | $^\ddagger$ | $^\ddagger$ |
| **Downvote** | .12 | $^\circ$ | $^\circ$ | .22 | $^\circ$ | $^\dagger$ | .20 | $^\circ$ | $^\circ$ |
| **Post Score** | | | | | | | | | |
| **Time of View** | .15 | $^\circ$ | $^\circ$ | .05 | $^\circ$ | $^\circ$ | .11 | $^\circ$ | $^\circ$ |
| **Final Score** | .13 | $^\circ$ | $^\circ$ | .13 | $^\circ$ | $^\circ$ | .17 | $^\circ$ | $^\circ$ |
| **SRNΔ** | | | | | | | | | |
| **Upvote** | .18 | $^\circ$ | $^\circ$ | .05 | $^\circ$ | $^\circ$ | .35 | $^\circ$ | $^\ddagger$ |
| **Downvote** | .10 | $^\circ$ | $^\circ$ | .17 | $^\circ$ | $^\circ$ | .15 | $^\circ$ | $^\circ$ |

We also found a severe lack of browsing and voting variety. Participants seem more inclined to exert more effort when interacting with a more diverse range of content, indicated by a greater probability of browsing further down and onto the second page when viewing a frontpage over a specific subreddit. However, the scope within a user's browsing session remains relatively narrow despite having access to a potentially wide range of diverse content and communities through the numerous subreddits available. Despite an intuitive increase in the number of subreddits as session length increases, users are typically engaging with a limited number of communities regardless of session length. There is a similar lack of variety in voting by users; most subreddits received very few votes and most voters voted in very few subreddits. This shows that while a diverse range of communities are available, users tend to self-select a narrow scope to interact with.

# Acknowledgments

This research is sponsored by the Air Force Office of Scientific Research FA9550-15-1-0003. The research was approved by University of Notre Dame Institution Review Board and the United States Air Force Surgeon General's Research Compliance Office. Reddit, Inc. was not involved in the experimental design, implementation, or data analysis.